\documentclass[a4paper,12pt,oneside]{article}
\usepackage[koi8-r]{inputenc}
\usepackage{graphicx}
\usepackage{amsmath,amssymb}			

\begin{document}

\title{Orbital motions and light curves of young binaries XZ~Tau and VY~Tau}

\author{A.~V.~Dodin$^1$, N.~V.~Emelyanov$^1$,  A.~V.~Zharova$^1$, S.~A.~Lamzin$^1$ 
\\
E.~V.~Malogolovets$^2,$ J.~M.~Roe$^3$}
\date{ \it \small
  $^1$ Sternberg Astronomical Institute, Moscow M.V. Lomonosov State University,
 Universitetskij pr., 13, 119992 Moscow, Russia\footnote{E-mail: lamzin@sai.msu.ru}\\
  $^2$ Special Astrophysical Observatory, N. Arkhyz, Karachai-Cherkesia 369167, Russia\\
  $^3$ AAVSO, PO Box 174, Bourbon, MO 65441, USA}

\maketitle
\bigskip


\bigskip

\begin{abstract}
  The results of our speckle interferometric observations of young 
binaries VY~Tau and XZ~Tau are presented. For the first time, we found 
a relative displacement of VY~Tau components as well as a preliminary 
orbit for XZ~Tau. It appeared that the orbit is appreciably non-circular 
and is inclined by $i \lesssim 47^o$ from the plane of the sky. It means 
that the rotation axis of XZ~Tau~A  and the axis of its jet are significantly 
non-perpendicular to the orbital plane. We found that the average brightness 
of XZ~Tau had been increasing from the beginning of the last century up to 
the mid-thirties and then it decreased by $\Delta B > 2$ mag. The maximal 
brightness has been reached significantly later on the time of periastron 
passage. The total brightness of XZ~Tau's components varied in a non-regular 
way from 1970 to 1985 when eruptions of hot gas from XZ~Tau~A presumably 
had occurred. In the early nineties the variations became regular following 
which a chaotic variability had renewed.  We also report that a flare activity 
of VY~Tau has resumed after 40 yr pause, parameters of the previous and new 
flares are similar, and the flares are related with the A component.
\end{abstract}

\section*{Introduction}

The most obvious manifestations of so-called `eruptive phenomena in 
early stellar evolution' (Herbig, 1977) are relatively short (from days 
to hundreds years) episodes of increased brightness observed in T Tauri 
stars, i.e. low-mass $(<2-3$~M$_\odot)$ young (age $<10$~Myr) stars that 
contracting toward the main-sequence.  It is widely accepted now that such 
events occur due to clearing of circumstellar gas-dust envelope or/and 
increasing of accretion rate $\dot M_{\rm acc}$ of matter from a protoplanetary 
disc (see, e.g. Audard et~al. (2014) and references therein). The physical reasons 
that are responsible for the $\dot M_{\rm acc}$ increasing are of special interest. 
In young binary systems such events can occur due to gravitational influence 
of each component of a system on the circumstellar disc of the neighboring 
star near the periastron passage. Such possibility was initially predicted 
theoretically (Bonnell, Bastien, 1992; Artymowicz, Lubow, 1996) and later 
observed in a number of young binaries (see, e.g.  Jensen et al. (2007) and references therein). 
The aim of our paper is to search for a correlation between the orbital motion and the photometric 
behavior in young binaries VY~Tau and XZ~Tau.

The photometric variability of XZ~Tau was discovered by P.F.Shain (Beljawsky, 1928). 
The star is projected onto the dark cloud L1551, has M3 spectral type, and its 
spectrum exhibits strong emission lines with equivalent width of H$_\alpha$ line 
up to 270~\AA.  The set of these features was the reason to include the star in 
the initial list of T Tauri stars (Joy, 1945) and to classify it from the present-day 
viewpoint as a classical T Tauri star (CTTS), i.e. low mass young star activity 
of which is due to accretion of protoplanetary disk's matter.  Haas et al. (1990) 
found that to north of the more bright (at that moment) star XZ~Tau~S there is 
a companion XZ~Tau~N at a distance of about 0.3~arcsec, what respects to projected 
distance of 42~AU if to adopt the distance to XZ~Tau to be 140~pc (Elias, 1978).  
It was found that both stars are CTTSs, and the N component has more early spectral 
type (M1.5) than the S component and therefore is more massive (White, Ghez, 2001).  
None the less it is adopted in the literature to designate XZ~Tau~S as the main 
component of the system, i.e.  as XZ~Tau~A, and XZ Tau N as a companion, 
i.e.  XZ~Tau~B.  We will also follow this tradition.

   XZ~Tau was identified by Mundt et al. (1990) as the source of a bipolar,
collimated outflow.  It appeared that the outflow occurs from each component of
the binary, and that the jet of XZ~Tau~A is oriented along a position angle
($\theta$) of $\sim 15^o$ while the jet of XZ~Tau~B along $\theta$ of $\sim 36^o$
(Krist et al., 2008).  Besides Krist et~al. (1997) discovered in the inner parts of the
outflow an unusual expanding bubble of emission nebulosity.  The bubble appears 
to be a succession of bubbles, the outer boundary of which reached an
extent of $\approx 6$~arcsec from the binary in 2005. The jet from XZ~Tau~A
is aligned with the outflow axis of the bubbles, and Krist et al. (2008) believe
that the bubbles are resulted from large velocity pulses of this jet.

Judging from the dynamical time of the bubble's expansion one can conclude 
that the large velocity pulses have occurred between 1965 and 1985.  Bear in
mind that high velocity ejection of gas in young binary DF~Tau was
accompanied with a short-time increase of stellar brightness of $\Delta B
\approx 6.5$~mag (Li et al., 2001), it would be interesting to search for similar
flares in XZ~Tau in the respective epoch.  Since the distance between
XZ~Tau's components gradually decreased after 1990, Krist et al. (2008)
concluded that the gas ejection was not connected with the moment of periastron
passage.  But Carrasco-Gonzalez et al. (2009) found from their Very Large Array (VLA)
observations that at 7~mm the XZ~Tau~A radio component resolves
into a binary with 0.09~arcsec (13~AU) separation, suggesting that XZ~Tau is
actually a triple star system.  They have assumed that the ejection of gas
from the XZ~Tau system may be related to a periastron passage of this newly
discovered close binary system XZ~Tau~AC.

  But follow-up VLA-observations of Forgan et al. (2014), as well as recent
observations of ALMA Partnership et~al. (2015) find no evidence of XZ~Tau~C component, despite
improved sensitivity and resolution.  Three possible interpretations were
offered: either XZ~Tau~C is transiting XZ~Tau~A, or the emission seen in
2004 by Carrasco-Gonzalez et al. (2009) was that of a transient, or XZ~Tau~C does not exist. 
Thus the question is open and it is a good time to analyze an additional
astrometric and photometric information about the system.

  VY~Tau, the second target of our investigation, is also very
interesting object.  Its photometric variability was discovered by
Beljawsky (1924), who reported that the photographic magnitude of the star varied
from 13 to 9.7~mag in 1906-1922.  Follow-up observations revealed
VY~Tau's light curve, featuring sporadic outbursts from about $B=14$ to
11-12~mag with duration from 100 to 700~days interspersed with many years of
inactivity (see Herbig (1990) and references therein).

  The star is projected onto the dark cloud L1536, has (at minimum light)
M0\,V spectral type with H$\alpha$ and Ca\,{\sc ii}~H,~K lines in emission
and strong Li\,{\sc i}~6707~\AA\, line in absorption.  Based on these
features (Herbig, Kameswara~Rao, 1972) have attributed VY~Tau to T Tauri type stars. 
The relatively small equivalent width of H$\alpha$ line in a quiet state
(EW$\approx 5$~\AA\,) and the negligible near-infrared (NIR) excess emission are
the reasons to conclude that the star has no accretion disk (Herbig, 1990) and
thus belongs to so-called weak-lined T Tauri stars (WTTS).

  According to the modern paradigm, an observed activity of WTTSs is attributed
to the presence of powerful chromospheres and coronas more or less similar to
solar.  But it is impossible to explain sporadic outbursts of VY~Tau using 
the analogy with solar flares.  At first, the duration of the outbursts is
almost three order of magnitudes larger than that of solar chromospheric
flares.  Secondly, while during the outburst VY~Tau becomes more blue,
low-level emission lines of Fe\,{\sc i}, Si\,{\sc i} and Mg\,{\sc i}
dominate in its spectrum, whereas higher excitation species such as Fe\,{\sc
ii} and the Balmer lines are relatively weak (Herbig, 1990), as contrasted to
solar flares.  Herbig suggested that the outburst activity of VY~Tau is
connected in one way or another with binarity of the star.  He also noted
that his analysis of VY~Tau spectrum was based on observations carried out
prior to 1972, when the star was last active, and there is need for a modern
reinvestigation particularly when the activity resumes.  We will see below
that the inactive state of the star lasted up to the end of 2012 following which
outbursts apparently have renewed.

  Leinert et al. (1993) discovered at a distance of about 0.7~arcsec a companion
VY~Tau~B, which was at that moment four times weaker in $K$-band than the
main star VY~Tau~A.  Follow-up observations of White, Ghez (2001)  and
McCabe et al. (2006) have confirmed the existence of the companion, but it was
impossible to state that the relative position of A and B components varies with
time.  Adopting the distance to VY~Tau to be 160 pc (Herczeg, Hillenbrand, 2014), the
projected distance between the components is $>100$~AU, so it looks unrealistic
to explain VY~Tau outburst activity as a result of interaction between A and
B components of the binary.  Therefore, if Herbig's hypothesis is correct,
then one more component should be in the system.  Thus -- as in the case of
XZ~Tau -- it is very important to obtain new photometric and astrometric
information about VY~Tau.

%
\section*{Observations and data reduction} \label{observations-sect}

Speckle interferometry of VY~Tau and XZ~Tau was performed with the 6-m
telescope of the Special Astrophysical Observatory of the Russian Academy of
Sciences on the night of 2014 February 12/13.  A system of speckle image
detection based on a CCD with internal signal amplification was used
(Maksimov et al., 2009).  Speckle interferograms were obtained with filters having
central wavelengths/half-widths of 700/40, 800/100, 900/60 nm in the case of
XZ~Tau and 800/100 nm in the case of VY~Tau.  2000 speckle images were
accumulated with exposure times of 50~ms per frame in each filter, to
determine the position parameters and magnitude differences.  The seeing was
$\approx 1.5-2$~arcsec but we observed both binaries through the haze and
the atmospheric extinction was $\sim 3$~mag.  As a result we were able to
accumulate the signal in the power spectrum not more than $\sim 30$ per cent of
the maximum possible level.  It practically did not decrease the accuracy of
angular distance $\rho$ and $\theta$  measurements, but reduced significantly
that of magnitude difference $\Delta m = m_B - m_A.$ All the more our data
are not suitable to search for the presumably faint close companion XZ~Tau~C
(Carrasco-Gonzalez et al., 2009). Results of our measurements and their errors are presented in
Table~\ref{bta-results}.  As can be seen from the table $\rho$ and $\theta$ 
values for XZ~Tau derived in different bands agree within errors of
measurements, and so we will use later on their averaged values.


\begin{table}
\caption{Results of our speckle interferometric observations}
\label{bta-results}
\begin{tabular}{@{}lcccccccc}
\hline
Set & Star & $\lambda^a$ & $\rho$ & $\sigma_{\rho}$ & $\theta$ & $\sigma_{\theta}$ 
& $\Delta m$ & $\sigma_m$ \\
 & Tau & $\mu$m & mas & mas & $^o$ & $^o$ &  &  \\
\hline
Feb. & VY & 0.8 & 712 & 8 & 323.9 & 0.7 & 1.8  & 0.2  \\
     & XZ & 0.7 & 268 & 8 & 311.6 & 1.6 & 0.0  & 0.5  \\
     &    & 0.8 & 269 & 3 & 311.4 & 0.6 & 0.0  & 0.5  \\
     &    & 0.9 & 272 & 5 & 311.6 & 1.1 & 0.0  & 0.5  \\
Apr. & XZ & 0.8 & 268 & 4 & 311.4 & 0.9 & 0.37 & 0.35 \\
     & VY & 0.8 & 714 & 7 & 323.5 & 0.6 & 1.84 & 0.15 \\
\hline
\end{tabular}\\
${}^a$: Central wavelength of a filter.
\end{table}


  We observed both binaries once more on night 2014 April 11/12 but with
800/100 nm filter only.  4000 speckle images were accumulated with exposure
times of 50~ms.  As can be seen from Table~\ref{bta-results} the results of
April and February observations coincide within errors of measurements. 
Unfortunately weather conditions were as bad as in February, so both
datasets cannot help to answer the question: if XZ~Tau~C companion exists
or not.

  The extensive photographic plate collection of the Sternberg Astronomical
Institute (SAI) was used to study the photometric behavior of XZ~Tau.  We found
262 plates with images of the vicinity of XZ~Tau.  260 of them were taken with the
40-cm astrograph of the Crimean Laboratory of SAI.  The plates cover a field
of $10^o\times10^o$ and have limiting magnitude $B \approx 17.$ The
photographic estimates were calibrated to stars USNO-A2.0.  The typical
uncertainty of our estimates is 0.3~mag.  Two more plates were obtained at
Moscow observatory of SAI in 1911-1912 by means of the 10-cm Steinheil astrograph.  
Both plates have a limiting magnitude of $B \approx 14.$
XZ~Tau is hardly distinguishable at these plates and we adopted an upper limit
13.5~mag to its brightness.

We also carried out 28 observations of VY Tau between March 26, 2010 and 
October 1, 2014 with the 35-cm Schmidt-Cassegrain telescope (amateur observatory, Bourbon, USA) 
equiped with SBIG STL-1001E/ST-402 CCD camera and a standard Johnson {\it V} filter. 
The CCD images were reduced in a standard way which included dark-frame 
and flat-field corrections. The image scale was $1.29-2.57$ arcsec per pixel, 
depending on the setup, while the size of stellar images was $FWHM\approx7-17$ 
arcsec. Some of these images are taken during the outbursts that allows 
us to determine which of two components increases its brightness by comparison 
of positions of the photo-centre of the system before/after and during the outburst. 
To measure the position of the photo-centre, 16 reference stars with small proper 
motion were selected from the USNO-B1.0 Catalog (Monet et al., 2003) in the field of 
$10\times10$ arcmin around the VY Tau.  We used these 16 stars in order to 
construct the coordinate system, assuming a linear transformation between the 
standard and image coordinates (Zombeck, 2007). Since the same stars were used for 
all obtained images, then possible systematic errors of our coordinate system 
must be common for all images and cancel out in the position difference measurements. 
The position of the photo-centre was determined in the image coordinates as a weighted 
mean $(i_c,j_c)=\sum_{i,j} (i,j)C_{ij}/\sum_{i,j}C_{ij}$ over a small square region, 
covering the star. Here $C_{ij}$ is the number of (sky-subtracted) flux counts in 
pixel $(i,j).$ Then the image coordinates were transformed to the right ascension 
$\alpha$ and declination $\delta$ of our coordinate system. For convenience we will 
use the following coordinates expressed in arcsec: 
$x = ({\alpha- \alpha_0})\cos \delta_0, y = {\rm \delta-\delta_0},$ 
where ${\alpha_0}$ and ${\delta_0}$ are mean coordinates of the star over all frames. 
Variations in the position and size of the region allow us to estimate an accuracy 
of determining the photo-centre $\sigma^*_{x,y}$. Quality of the constructed coordinate 
system is characterized by mean deviations $\sigma^c_{x,y}$ of positions of five control 
stars from their average positions calculated over all frames. The control stars are stars 
of 12-14 mag, for which shifts of the coordinates due to proper motions and parallaxes 
can be neglected during the period of our observations. The total uncertainty of the 
coordinates of the VY Tau in our system can be estimated as 
$\sigma_{x,y} = \sqrt{\sigma^{*2}_{x,y} + \sigma^{c2}_{x,y}}.$ 
Results of the astrometric and photometric measurements are summarized in Table~\ref{vy}.

\begin{table}
 \caption{Photometric and astrometric measurements of the VY Tau}
  \label{vy}
\begin{tabular}{@{}lcccccc}
\hline
JD2456...&  $V$  & $\sigma_V$ &   $x$  &   $\sigma_x$&   $y$  &  $\sigma_y$\\
         &  mag   &    mag & $ ^{\prime\prime}$ &  $ ^{\prime\prime}$ &$ ^{\prime\prime}$ & $ ^{\prime\prime}$ \\
\hline
  012.63 &  13.59 &   0.01 &   -0.11 &    0.27 &    0.02 &    0.16 \\ 
  012.63 &  13.59 &   0.01 &    0.01 &    0.09 &   -0.14 &    0.13 \\ 
  012.63 &  13.60 &   0.01 &   -0.08 &    0.29 &   -0.04 &    0.12 \\ 
  013.61 &  13.59 &   0.06 &    0.46 &    0.21 &    0.40 &    0.42 \\ 
  147.91 &  13.84 &   0.09 &    0.03 &    0.52 &   -0.06 &    0.17 \\ 
  247.84 &  13.25 &   0.05 &   -0.16 &    0.08 &    0.21 &    0.36 \\ 
  251.85 &  12.98 &   0.03 &   -0.02 &    0.16 &   -0.06 &    0.31 \\ 
  253.84 &  13.00 &   0.03 &    0.13 &    0.16 &    0.02 &    0.22 \\ 
  267.83 &  12.58 &   0.07 &    0.06 &    0.16 &    0.09 &    0.11 \\ 
  280.83 &  12.79 &   0.01 &   -0.01 &    0.09 &   -0.04 &    0.13 \\ 
  299.82 &  13.29 &   0.05 &   -0.00 &    0.13 &   -0.00 &    0.14 \\ 
  328.75 &  13.61 &   0.07 &   -0.22 &    0.33 &   -0.28 &    0.17 \\ 
  574.96 &  14.00 &   0.02 &    0.05 &    0.21 &   -0.22 &    0.29 \\ 
  583.94 &  13.75 &   0.02 &   -0.54 &    0.41 &   -0.05 &    0.24 \\ 
  590.90 &  13.98 &   0.02 &    0.03 &    0.36 &    0.01 &    0.31 \\ 
  593.89 &  13.88 &   0.02 &   -0.32 &    0.21 &    0.24 &    0.20 \\ 
  599.89 &  13.77 &   0.02 &   -0.07 &    0.37 &    0.18 &    0.35 \\ 
  604.96 &  13.32 &   0.01 &    0.07 &    0.19 &    0.03 &    0.16 \\ 
  606.93 &  13.41 &   0.01 &    0.01 &    0.17 &   -0.01 &    0.14 \\ 
  609.90 &  13.10 &   0.01 &   -0.14 &    0.15 &   -0.10 &    0.07 \\ 
  626.86 &  12.85 &   0.02 &    0.05 &    0.14 &    0.02 &    0.12 \\ 
  639.79 &  12.70 &   0.03 &    0.18 &    0.33 &   -0.17 &    0.10 \\ 
  653.82 &  13.04 &   0.02 &    0.09 &    0.15 &   -0.05 &    0.06 \\ 
  894.84 &  13.82 &   0.03 &    0.37 &    0.18 &    0.04 &    0.25 \\ 
  895.83 &  13.87 &   0.04 &   -0.06 &    0.19 &   -0.11 &    0.16 \\ 
  905.80 &  13.82 &   0.06 &    0.33 &    0.30 &   -0.02 &    0.09 \\ 
  930.83 &  13.83 &   0.03 &    0.05 &    0.15 &   -0.02 &    0.08 \\ 
  931.73 &  13.79 &   0.07 &   -0.20 &    0.23 &    0.13 &    0.16 \\ 
\hline
   \end{tabular}
  \end{table}


%
\section*{Orbital motion} \label{orbit-sect}

 For both binaries we searched the literature for previously published
measurements of their separations.  These data are presented in
Table~\ref{position} along with our data averaged over different filters in
the case of February data for XZ~Tau.  Note that to map the visual orbit of
XZ~Tau we use measurements obtained in optical and NIR bands but not in
radio band because it is not obvious that centroids of radio emission
coincide with components of the system (see Sect.~\ref{discussion}).


\begin{table}
 \caption{Compilation of XZ~Tau and VY Tau measurements}
  \label{position}
\begin{tabular}{@{}lcccccc}
\hline
Star & JD24...& $\rho$ & $\sigma_{\rho}$ & $\theta$ & $\sigma_{\theta}$ & Ref. \\
     &        & mas    &  mas            & $^o$ & $^o$      &      \\
\hline
XZ Tau  & 47817 & 300   & 20   &  334   &  3   & 1 \\
        & 48550 & 311   &  6   &  333   &  2   & 2 \\
        & 49723 & 306   &  3   &  327.2 &  0.5 & 7 \\
        & 50423 & 299.8 &  5.7 &  324.5 &  1.0 & 3 \\
        & 50516 & 300   &  3   &  326.0 &  0.5 & 7 \\
        & 51213 & 301   &  3   &  324.4 &  0.5 & 7 \\
        & 51581 & 299   &  3   &  322.6 &  0.5 & 7 \\
        & 51881 & 299   & 14   &  322.6 &  2.0 & 4 \\
        & 51951 & 297   &  3   &  321.6 &  0.5 & 7 \\
        & 52318 & 291   &  3   &  321.6 &  0.5 & 7 \\
        & 53025 & 294   &  3   &  322.3 &  0.5 & 7 \\
        & 56703 & 270   &  3   &  311.5 &  0.7 & 8 \\
        & 56759 & 268   &  4   &  311.4 &  0.9 & 8 \\
VY Tau  & 48231 & 660   & 20   &  317   &  2   & 6 \\
        & 50792 & 665   & 13   &  316.6 &  1.0 & 3 \\
        & 52245 & 670   & 10   &  318   &  1   & 5 \\
        & 56703 & 712   &  8   &  323.9 &  0.7 & 8 \\
        & 56759 & 714   &  7   &  323.5 &  0.6 & 8 \\
\hline
   \end{tabular}\\
1 -- Haas et al. (1990); 2 -- Ghez et al. (1993); \\
3 -- White, Ghez (2001); 4 -- Hartigan, Kenyon, (2003) \\
5 -- McCabe et al. (2006); 6 -- Leinert et al. (1993); \\
7 -- Krist et al. (2008); 8 -- our data.
  \end{table}

\begin{figure} \includegraphics[width=150mm]{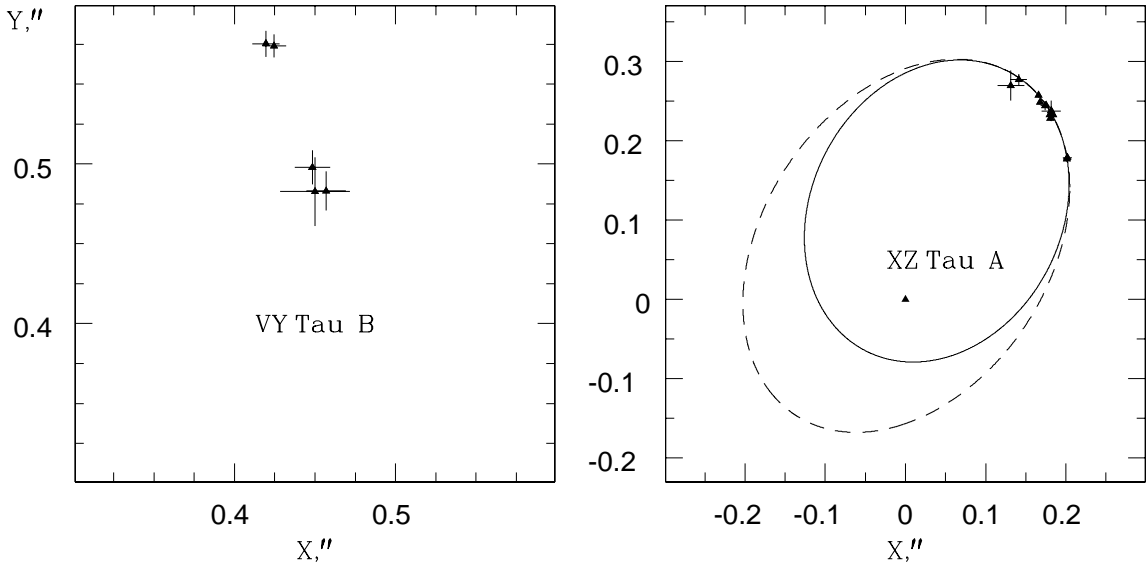}
\caption{Variations of relative position of components in VY~Tau (left
panel) and XZ~Tau (right panel)-- see Tabl.~\ref{position}.  Two orbits from the family of
Table~\ref{orbit-param} are shown for XZ~Tau: with $e=0.6$ (solid line) and
with $e=0.3$ (dashed line).  See text for details.  In both panels the
origin of a coordinate system is in the A component, north is up and east is to
the left.} \label{fig:period} \end{figure}

  Variation of relative position of VY~Tau and XZ~Tau components based on
data presented in Table~\ref{position} is shown in Fig.~\ref{fig:period}.  In
the case of VY~Tau our measurements for the first time indicate
the shift of the companion relative to the main star although its motion is
indistinguishable from linear now.  If the system is bounded, then its
orbital period is $>350$~yr that follows from the third Kepler's law, if to
adopt the following parameters of the system: the distance to VY~Tau is 
160~pc (Herczeg, Hillenbrand, 2014); the semi-major axis $a$ is larger than a half of observed
distance between the components $(\approx 0.7$~arcsec) i.e.  $a>56$~AU; the total
mass of the system $M_A+M_B < 1.2$~M$_\odot$ (Woitas et al., 2001).

 In the case of XZ~Tau there is enough orbital coverage to compute
preliminary orbital solutions.  As far as an information on the existence of a
third body in the system is absent, Kepler's two-body problem is considered. 
We calculated orbital elements in a Cartesian coordinate system with the
origin in the A component, Z-axis directed to the Earth and X-axis directed
to the north celestial pole of J2000 epoch.  The distance to XZ~Tau was
adopted to be 140~pc (Elias, 1978; Herczeg, Hillenbrand, 2014).


  The orbital elements were found from the observational data given in
Table~\ref{position} by differential corrections calculated with the
least-squares method.  Weights were applied using the accuracy estimates of
observations.  The total mass of the system was treated as a parameter of
the problem therefore values of semi-major axis and mean motion were
considered as independent variables.


   It appeared that the available observational data do not allow to
determine all orbital elements jointly due to strong correlations between
them, arising from the least-squares method process.  Some family of
orbits represents the observations with almost equal accuracy.

  For this reason we used inclination of the orbital plane from the plane of
the sky $i$ as an independent parameter $(0 < i < 90^o).$ For a set of
$i$-values we used an eccentricity $e$ as the second free parameter and
found that inside some range of its value the process of differential
correction converges and one can derive the rest of the orbital elements.

  One can impose restrictions on the total mass of the binary bearing in mind
that the spectral type of each component is M.  According to Hartigan, Kenyon (2003) and
Herczeg, Hillenbrand (2014) the mass of XZ~Tau~A is $M_A=0.29$~M$_\odot.$ An estimation of
mass of XZ~Tau~B is not so reliable because its spectrum is strongly
contaminated by emission lines: Hartigan, Kenyon (2003) found $M_B=0.37$~M$_\odot,$
whereas White, Ghez (2001) found $M_B=0.51 \pm 0.08$~M$_\odot.$ Thus we adopted
the total mass of XZ~Tau as $M_A+M_B < 0.9$~M$_\odot.$ It appeared that
only orbits with $i\le 47 ^o$ satisfy this restriction.  The allowable
range of orbital parameters of interest (eccentricity, semi-major axis,
orbital period $P,$ time of periastron passage $T_0$ and total mass of the
binary $M_{AB})$ for a set of $i$-values are presented in
Table~\ref{orbit-param}.  The last row of the table is presented for
comparison.

\begin{table*}
\begin{minipage}{126mm}
\caption{Allowable range of XZ~Tau orbital elements}
  \label{orbit-param}
\begin{tabular}{l c c c c c }
\hline
$i$      & $e$ & $a$ & $P$ & $T_0$ & $M_{AB}$  \\
 $^o$ &     &  AU &  yr &  yr   & M$_\odot$  \\
\hline
10 &  0.62$-$0.64 & 27.2$-$27.3 & 155$-$156 & 1915.9$-$1916.4 & 0.82$-$0.84 \\
20 &  0.58$-$0.64 & 28.3$-$28.4 & 158$-$167 & 1915.3$-$1919.4 & 0.80$-$0.90 \\
30 &  0.52$-$0.56 & 28.5$-$28.9 & 159$-$169 & 1919.4$-$1923.4 & 0.84$-$0.90 \\
35 &  0.45$-$0.56 & 31.4$-$32.0 & 190$-$197 & 1910.3$-$1913.3 & 0.80$-$0.90 \\
40 &  0.38$-$0.52 & 33.3$-$34.4 & 214$-$219 & 1904.6$-$1907.5 & 0.79$-$0.90 \\
45 &  0.29$-$0.48 & 36.1$-$38.2 & 241$-$256 & 1893.8$-$1899.3 & 0.80$-$0.90 \\
50 &  0.27$-$0.84 & 32.2$-$37.3 & 124$-$206 & 1915.5$-$1943.0 & 0.99$-$2.99 \\
\hline
  \end{tabular}
 \end{minipage}
\end{table*}

  We conclude therefore that the parameters of XZ~Tau preliminary orbit can be
described as follows: $i<47 ^o,$ $0.29 < e < 0.64,$ $27.2 <a < 37.2$~AU,
$155 < P < 256$ yr.  These values can be compared with orbital parameters of
the binary published earlier: $i\approx 0^o,$ $e \sim 0.9,$ $a>21$~AU,
$P>99$~yr (Krist et al., 2008); $i=17^o,$ $e\sim 0.5 \pm 0.2,$ $a=84$~AU, $P=1010
\pm 260$~yr (Hioki et~al., 2009).

  Two orbits from the family of Table~\ref{orbit-param} with significantly
different eccentricities are shown in the right panel of
Fig.~\ref{vytau-light-curve} to demonstrate that in both cases theory
describes observational data equally good.  Parameters of these orbits are:
$e=0.6,$ $i=20^o,$ $a=28.4$~AU, $P=166$~yr and $e=0.3,$ $i=45^o,$
$a=36.4$~AU, $P=244$~yr.

%
\section*{Discussion} \label{discussion}

\begin{figure}
\includegraphics[width=150mm]{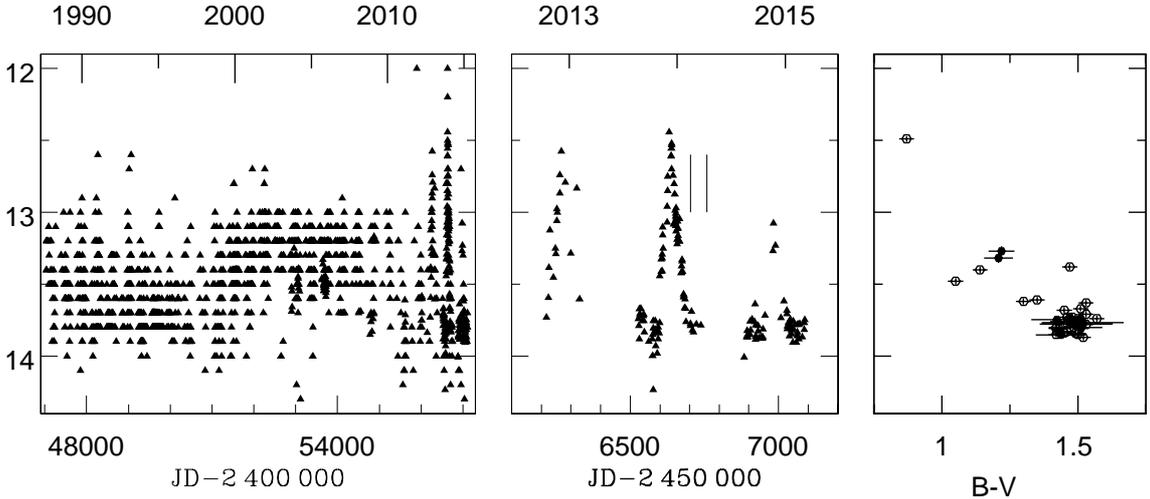}
\caption{
Left panel: $V$-band light curve of VY~Tau from 1988 to 2015 based on
visual and photoelectric AAVSO observations.  Middle panel: a zoomed
fragment of the same light curve showing renewed outbursts in more detail,
only photoelectric measurements (including data from Tabl.~\ref{vy}) are plotted.  
Vertical lines mark
moments of our speckle interferometric observations.  Right panel: variations
of $V$-magnitude of the star vs color index $B-V.$ Filled circles represent the
data observed after 2012 (the AAVSO database) and open circles are for the old
photoelectric data adopted from the literature (Stone, 1983;
Rydgren et al., 1984; Kolotilov, 2001; Grankin et al., 2007).
} 
\label{vytau-light-curve} 
\end{figure}

 The orbital period of VY~Tau~AB binary is $>350$~yr therefore it seems
unreasonable to expect a direct connection between the orbital motion of the
binary components and the non-regular outbursts observed in VY~Tau in the last
century.  As was noted in the Introduction, the outburst activity has terminated
at the beginning of 1970s.  It can be seen from Fig.~\ref{vytau-light-curve}
based on AAVSO data (Kafka, 2015) that the inactive period lasted almost up to
2013 i.e.  $\approx 40$~yr\footnote{
{\it Short} $(\Delta t <1^h)$ flares with amplitudes $\Delta U$
up to 1.5~mag were observed by Khodzhaev (1987) in 1983.
}
and then the outbursts renewed.

  At the date of writing, three new outbursts have been observed. Their
amplitude $\Delta V$ reaches 1.5~mag, duration is $\sim 100^d$ and the time
interval between them is $\approx 1$~yr. These parameters look like the parameters
of outbursts observed in the last century.  We found 9 nearly
simultaneous observations of VY~Tau in $B$ and $V$ filters in the AAVSO database
after 2012, two of which were carried out at the epoch of increased
brightness.  Judging by these data, color indexes $B-V$ of the star in active
and quiet states are also similar to that of previous -- see the right panel
of Fig.~\ref{vytau-light-curve}.

\begin{figure}
\includegraphics[width=84mm]{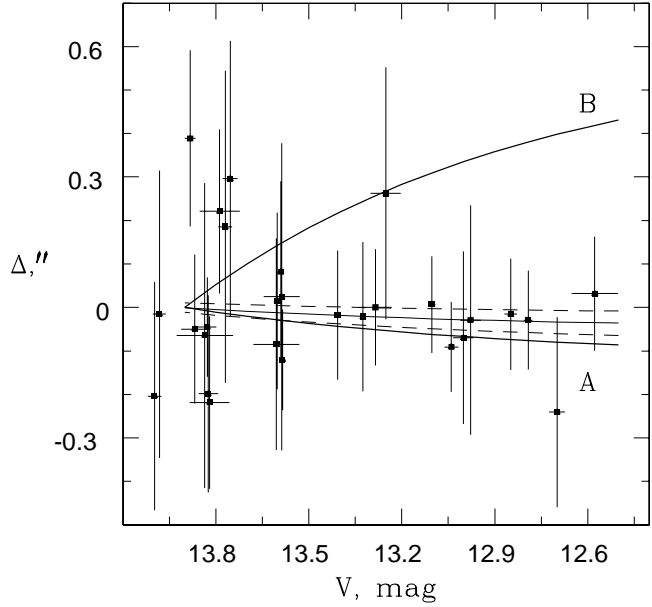}
\caption{The shift of VY Tau's photo-centre as a function of V magnitude. Dots with 
error bars are for the observations. Thick solid curves labelled as A and B are for 
the expected shifts with $r_0=5$ in the case when the outbursts occur on the A or B 
component, correspondingly. If $r_0>5,$ then the slope of the A-curve will be lower, 
while the slope of the B-curve will be steeper according to Eq.~(\ref{eq1}). A linear 
regression estimated by weighted least squares and its 1-$\sigma$ uncertainties are 
shown by thin solid and dashed lines, respectively.} 
\label{vytauAstr} 
\end{figure}

 It has not been known up to now which component of VY Tau is responsible 
for the outbursts. Unfortunately, our speckle interferometric observations 
were carried out just after the end of the second outburst -- see the middle 
panel of Fig.~\ref{vytau-light-curve}. Nevertheless, it appeared possible to 
solve the problem by means of our unresolved astrometric and photometric 
measurements presented in Table~\ref{vy}. Using results of our speckle 
interferometric observations from Table \ref{bta-results}, one can find 
an unit vector ${\bf n} = (\sin\theta, \cos\theta)$ in the plane of the sky, 
directed from A to B. A shift of the photo-centre of the star along ${\bf n}$ 
can be found as $\Delta = x\sin\theta+y\cos\theta$ and its uncertainty 
$\sigma_\Delta = \sqrt{\sigma_x^2 \sin^2\theta + \sigma_y^2 \cos^2\theta}.$ 
Here $x,y,\sigma_x,\sigma_y$ are the position of the star with corresponding 
uncertainties from Table~\ref{vy}. The shift $\Delta$ is shown in Fig.~\ref{vytauAstr} 
as a function of the brightness of the star.

As a reference position of the star, we chose its position in a faint state. It can be 
easily shown that the shift of the photo-centre relative to the reference position along 
the straight line passing through A and B is
\begin{equation} \label{eq1}
   \Delta = \rho k\left[1-10^{0.4(V-V_0)}\right], 
\end{equation} 
where $\rho$ is the separation between the components; $V,$ $V_0$ denote the total 
$V$-magnitude of the binary during an outburst and in a quiet state, correspondingly. 
If the outbursts occur on the A component, then $k=k_{\rm A}=-(1+r_0)^{-1},$ while in the case 
of outbursts on the B component $k=k_{\rm B}=r_0(1+r_0)^{-1}.$ Here $r_0$ is the flux ratio of 
A and B components in the quiet state. Because the fainter component is cooler 
(Woitas et al., 2001), then the flux ratio in the $V$ band must be larger than at 800~nm, 
i.e. $r_0>5$ (Table~\ref{bta-results}) that means $-0.16<k_{\rm A}<0$ and $0.83<k_{\rm B}<1.$

We plot in Fig.~\ref{vytauAstr} the expected 
shifts $\Delta,$ calculated with $r_0=5,$ for the cases when the outbursts occur 
on either the A or B component. It can be seen that the observations do not show 
any significant shift in the direction of the B component during the outburst, 
therefore we can conclude that the outbursts occur on (or near) the A component. 
It follows from Eq.~(\ref{eq1}) that a higher value of $r_0$ can only strengthen 
our conclusion. We also found that there is no difference in the position of the 
VY Tau in active and quiet states in the direction orthogonal to ${\bf n}.$

 The source of the outbursts can be found in a more formal way.
It is possible to fit the observations by Eq.~(\ref{eq1}), 
re-written in a linear form $\Delta=\rho kt+b,$ where $t=1-10^{0.4(V-V_0)}$ and $b$ 
is a possible small shift due to an ambiguity of the definition of the reference position. 
We have found by weighted least squares (weights are $\sigma^{-2}_\Delta$) that 
$k=-(7\pm3)\times10^{-2},$ $b=(0.1\pm1.1)\times10^{-2}$ arcsec that results 
in $r_0=9-24$ and agrees well with our expectations of $r_0>5.$ 
One can definitely conclude now that the A component increases its brightness during 
the outbursts because the derived $k$-value differs from the low limit of 
$k_B$-value by 30 standard errors. These results are very important but do 
not reveal the nature of the outburst activity.


\begin{figure}
\includegraphics[width=150mm]{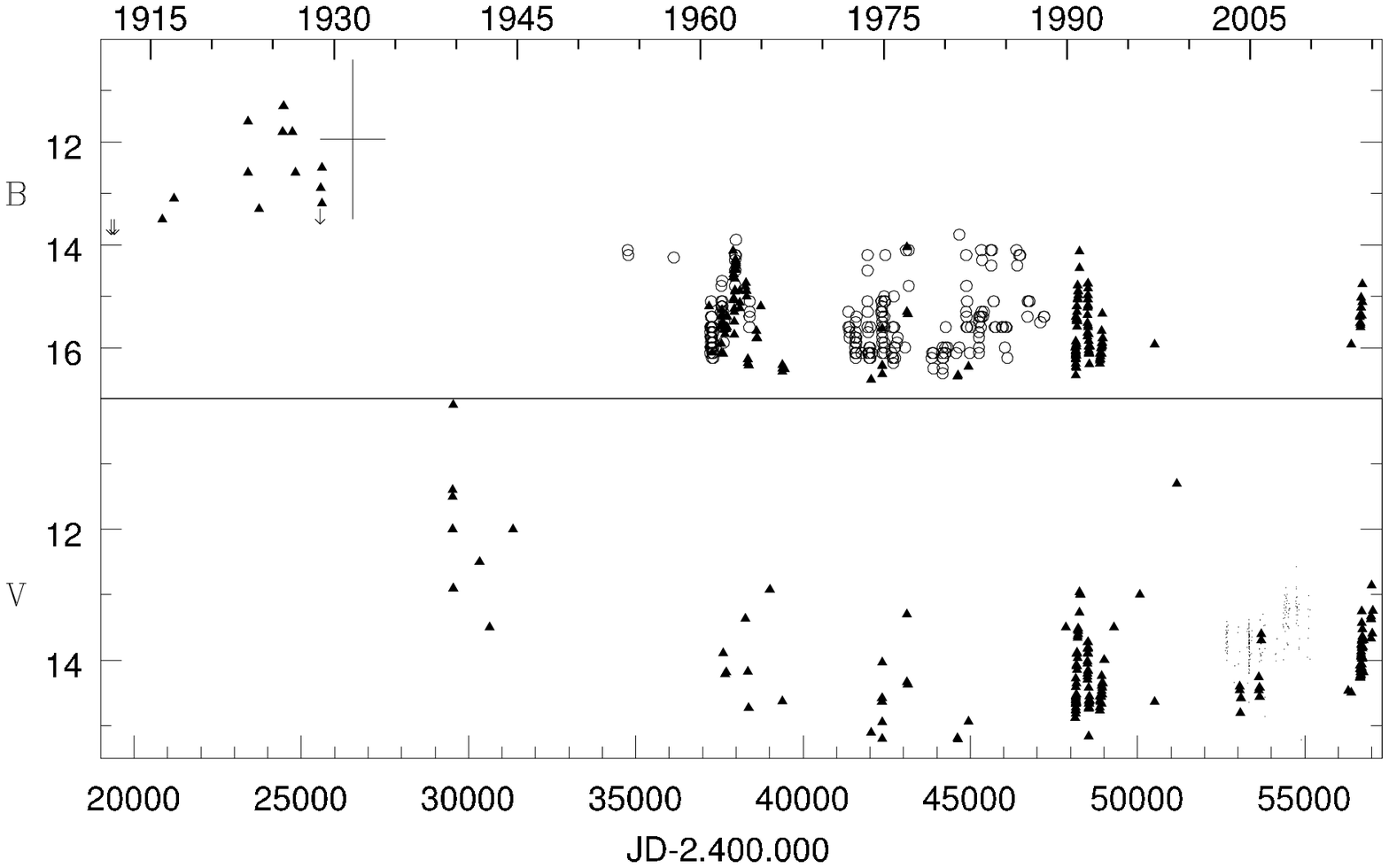}
  \caption{
XZ~Tau's light curve in $B$ (upper panel) and $V$ (low panel) bands.
The following designations are used: triangles are for the data adopted from the
literature, open circles are for our data, points are for the ASAS data,
arrows are upper limits. The large cross near JD~2.426.500 represents data
from R\"ugemer (1935) -- see text for details.
} 
 \label{xztau-light-curve}
\end{figure}

   XZ~Tau is also non-trivial object. We constructed the historical light curve
of the binary (see Fig.~\ref{xztau-light-curve}) using our data along with
data adopted from the literature (Beljawsky, 1928; Esch, 1929;
R\"ugemer, 1935; Joy, 1945; Smak, 1964; Badalyan, Erastova, 1964; Cohen, Schwartz, 1976;
Badalyan, Erastova, 1969; Rydgren, Vrba, 1981; Rydgren, Vrba, 1983; Rydgren et al., 1976;
Rydgren et al., 1982).  Unfortunately, R\"ugemer (1935) did not present results of his
109 estimations of XZ~Tau's brightness and wrote only that $B$-magnitude of
the star varied between 10.4 and 13.5 in JD~$2.425.528-2.427.513$ period.

  Additional data were taken from the following databases: the AAVSO
(Kafka, 2015), the T Tauri database (Herbst et al., 1994) and the ASAS project
(Pojmanski et al., 2005).  Note that $V$-magnitudes of XZ~Tau from the ASAS data could be
overestimated up to 0.5 due to a contribution from HL~Tau $(V\simeq 14.7$~mag
according to the AAVSO data) which is at the distance $\approx 20$~arcsec from
the binary.  As far as a separation between A and B components of XZ~Tau is
$\sim 0.3$~arcsec, all mentioned above data represent their summary brightness.

  As can be seen from Fig.~\ref{xztau-light-curve} the average brightness of
XZ~Tau increased from the beginning of 20th century, has reached the maximum {\it
circa} 1935 and then decreased significantly.  We suppose that observed
brightness variations are connected with variations of the accretion luminosity
of XZ~Tau~A and/or XZ~Tau~B components due to changing of the distance
between them.  It follows from Table~\ref{orbit-param} that the periastron 
passage has happened before 1924, i.e. significantly earlier than
the moment of XZ~Tau's maximal brightness.  But a companion produces the
most strong tidal perturbations in the outermost regions of a neighbouring
star's circumstellar disk, and some time $\Delta t$ is necessary for the
disturbance to propagate inward and to result in increasing of accretion rate
onto the central star. Thus it looks natural that the maximum of accretion
luminosity occurs later than the moment of a periastron passage. This time 
delay $\Delta t$ must be of order of Keplerian time at the outer boundary of 
the disk (Lamzin et al., 2001).

  According to Artymowicz, Lubow (1996) the outer radii $R_d$ of circumcomponent's disks
of a young binary is near 0.4 of a minimal distance between its components,
it means, in our case (see Table~\ref{orbit-param}), that $R_d\approx 6$~AU. 
The mass of XZ~Tau~A or B component is $\sim 0.5$~ M$_\odot$ (see Sect. 
\ref{orbit-sect}) and thus $\Delta t \sim 30$~yr.  Such value of the time
delay means that the time of periastron passage was {\it circa} 1905,
i.e. within the range of $T_0$-values in Table~\ref{orbit-param}.  We found
that the orbital period of the binary is $P=200 \pm 50$~yr therefore the maximum
of the XZ~Tau's brightness corresponds to the phase $\Delta t /P \sim
0.1-0.2$ of its orbital period.

  Note that in the case of another young binary DF~Tau even more significant
delay of a maximum of accretion luminosity relative to a time of a periastron
passage is observed.  The orbital period of the binary is $\approx 43.7 \pm
3.0$~yr (Schaefer et al., 2014) and that coincides with the duration of DF~Tau's photometric
cycle $(\approx 44$~yr according to Tsesevich et al., 1967) and also approximately
equals to a half of the time interval between two extremely strong flares
observed in DF~Tau in October 1918 and January 2000.  As was mentioned in
the Introduction (Li et al., 2001), a powerful gas ejection was observed from DF~Tau
during the second flare.

  According to Schaefer et al. (2014) the epoch of DF~Tau's periastron passage is
$1980.5 \pm 1.7$~yr, therefore the second flare has occurred at an orbital
phase $\approx 0.45,$ when the companion was near the apastron.  What is
more, DF~Tau's minimal accretion luminosity was observed in June 1987
(Lamzin et al., 2001), i.e.  at orbital phase $\sim 0.16.$ On the other hand,
it is stated that in the case of young binaries DQ~Tau 
(Mathieu et al., 1997), UZ~Tau~E (Jensen et al., 2007) and AK~Sco (Gomez~de~Castro et al., 2013)
the moments of maximal accretion luminosity and periastron passage coincide 
within errors of measurements. Note, just in case, that the orbital
periods of DQ~Tau, UZ~Tau~E and AK~Sco are of order of several tens of
days whereas that of XZ~Tau and DF~Tau are much more longer.

  As was mentioned in the Introduction, there are reasons to suppose that gas
ejections from XZ~Tau~A occurred between 1965 and 1985.  Our estimates of
XZ~Tau's brightness mainly cover the period from 1970 to 1985 -- see
Fig.~\ref{xztau-light-curve}.  During this period the brightness of the
binary varied non-regularly and sometimes the variations were rapid enough:
e.g.  $B$-magnitude of XZ~Tau was 16.0 in 1981, March 1 and $B=13.8$ five
days later.\footnote{
Duration of the powerful flare observed in DF~Tau in 2000 $(\Delta
B \approx 6.5$~mag) was 2-3 days only, so such short event in XZ~Tau can be
simply missed.
}
As can be seen from the figure, at the beginning of the nineties the
character of XZ~Tau's photometric variability has changed radically: smooth
wavy variations of XZ~Tau's brightness were observed during 1990-1993
(Grankin et al., 2007), but then a chaotic variability has renewed.  It is not clear,
however, if a chaotic variability of 1970-1985 was connected in some way
with the gas ejection because similar variability was also observed in XZ~Tau in
the beginning of the sixties.

   Note that it is not so obvious which component of the binary is
responsible for the rapid non-regular variations of XZ~Tau's brightness.  From
the beginning of the nineties, when resolved photometry of the components
became possible, XZ~Tau~B demonstrated larger photometric activity than
XZ~Tau~A and sometimes the B component becomes brighter than A, at least at
wavelengths $>0.8$~$\mu$m -- see Krist et al. (2008) and references therein. 
According to Table~\ref{bta-results} the brightness of A and B components
was the same within errors of measurements in 700/40 and 800/100~nm filters
during our February observations as well as in 800/100~nm filter two months
later.  On the other hand, from the beginning of the forties all authors who
determined the spectral type of XZ~Tau have found M3 value, which is the
spectral type of the A component, it apparently means that at the moments of
that (unresolved) spectral observations XZ~Tau~A was brighter in the optical
band than B.

  Note one more interesting feature of XZ~Tau. It looks natural to assume
that a bipolar gas outflow in {\it single} CTTS occurs in the direction that
is perpendicular to its accretion disk's midplane and (approximately) along
a rotation axis of the star.  The jet of XZ~Tau~A (Krist et al., 2008) as well as the
rotation axis of the star (Artemenko et al., 2012) are inclined by $63^o$ to the
line of sight.  On the other hand, we found that the inclination of the XZ~Tau
orbital plane to the plane of the sky is $<47^o,$ and therefore axes of
XZ~Tau~A rotation and jet makes an angle $>16^o$ with the normal to the
orbital plane probably due to gravitational influence of the B component.  It
seems useful to note in this connection that position angles of XZ~Tau~A
and XZ~Tau~B jets are different: $\approx 15^o$ and $\approx 36^o,$
respectively (Krist et al., 2008).

  One can assume also that jets of the components are not parallel to each
other and not perpendicular to the orbital plane, because there is a third
body in the system, the orbit of which is inclined significantly to the orbital
plane of XZ~Tau~AB binary.  However recent observations of ALMA Partnership et~al. (2015) as
well as previous VLA-observations Forgan et al. (2014) did not confirm the existence
of XZ~Tau~C companion.

   ALMA Partnership et~al. (2015) found from their observations at $\lambda=2.2$~mm the
following astrometric parameters of the binary: $\rho = 0.273 \pm 0.001$
arcsec, $\theta=308.7 \pm 0.5^o.$ Comparison with our data
(Table~\ref{bta-results}) indicates that their $\rho$-value is within the error
box of our measurements, but position angles in optical and radio bands
differ at more than $3\sigma.$ Does it really mean that centroids of optical
and radio emission have different position?  We have no answer to this
question, but one can not exclude such possibility.  It is why we did not
use results of very precise radio observations to map the orbit of XZ~Tau.

%
\section*{Concluding remarks} \label{conclusions-sect}

Our new astrometric and photometric data reveal a number of interesting
features in young binaries VY~Tau and XZ~Tau. Additional observational 
data are necessary in order to interpret this information in a proper way 
and we would like to mention in this Section which information seems the 
most crucial for us.

{\bf VY~Tau.} It is necessary to continue the photometric monitoring of 
the system because its unusual outburst activity apparently has renewed 
after a forty-year pause.  Spectral observations of the binary at different 
phases of future outbursts are necessary to reveal if they are connected with
chromospheric or accretion processes.  But resolved spectroscopy of the
binary in a quiet state is not less important, because we still
did not know even a spectral type of the B component.
If the outburst activity of VY~Tau is indeed connected 
with a binarity, as it was suggested by Herbig, then one can state now that 
the A component should be a close binary. Indeed, if the companion of VY~Tau~A 
is a CTTS, then it well can be that the outbursts are not connected with 
a chromospheric activity of the main (WTTS) star, but are a consequence 
of an accretion process in a hypothetical companion VY~Tau~C.

{\bf XZ Tau.} New high-precision astrometric observations will enable 5-10
years from now to improve significantly orbital elements of the binary, the
most interesting of which (from astrophysical viewpoint) are inclination
angle, eccentricity, orbital period and the time of a periastron passage. 
There is a deficit of photometric observations of the star during last 10
years, while AAVSO data indicate that large $(\Delta V > 1$ mag) variations
of XZ~Tau brightness have occurred. It would be interesting also to investigate
XZ~Tau's brightness variations, using old photographic plates of the Harvard
College Observatory Astronomical Plate Stack.  Multicolor resolved
photometry of XZ~Tau's components will help to understand which component is
responsible for flare activity in the optical band.  It is necessary to
clarify the nature of extended structures around A and B components of the
binary observed in radio band (Carrasco-Gonzalez et al., 2009; ALMA Partnership et~al., 2015).  Indeed,
these structures have almost round shape and if they are accretion disks
observed nearly pole-on, then the axes of jets are significantly
non-perpendicular to the disk's midplane that seems very unusual.

  And last but not least, for both binaries the question on the existence of
a third body in a system is equally important, but still open.

%
\section*{Acknowledgments}

We thank N.Samus and S.Antipin for useful discussions. We acknowledge with
thanks the variable star observations from the AAVSO International Database
contributed by observers worldwide and used in this research along with ASAS
project data.  This research has made use of the SIMBAD database and the
VizieR catalog access tool, operated at CDS, Strasbourg, France.  This work
was supported in part by the Program for Support of Leading Scientific
Schools NSh-261.2014.2 and NSh-2043.2014.2.

%

\end{document}